\documentclass{aa}

\usepackage{lscape}
\usepackage{txfonts}
\usepackage{graphicx}
\usepackage{rotating}
\usepackage{natbib}
\usepackage{gensymb}
\usepackage{url}
\usepackage{esvect}

\begin{document}

\title{Dissecting a SN impostor's circumstellar medium: MUSEing about the SHAPE of $\eta$~Car's outer ejecta\thanks{Based on observations collected at the European Organisation for Astronomical Research in the Southern Hemisphere under ESO programme 094.D-0215(A).}
} 


\author{A. Mehner\inst{1}
\and W. Steffen\inst{2}   
\and J.H. Groh\inst{3}
\and F.P.A. Vogt\inst{1}
\and D. Baade\inst{4}
\and H.M.J. Boffin\inst{1,4}
\and K. Davidson\inst{5}
\and W.J. de Wit\inst{1}
\and R.M. Humphreys\inst{5}
\and C. Martayan\inst{1}
\and R.D. Oudmaijer\inst{6}
\and T. Rivinius\inst{1}
\and F. Selman\inst{1}}

  \institute{ESO -- European Organisation for Astronomical Research in the Southern Hemisphere, Alonso de Cordova 3107, Vitacura, Santiago de Chile, Chile
  \and Instituto de Astronom\'{i}a, Universidad Nacional Aut\'{o}noma de M\'{e}xico, Apdo Postal 106, Ensenada 22800, Baja California, M\'{e}xico
 \and School of Physics, Trinity College Dublin, Dublin 2, Ireland
  \and ESO -- European Organisation for Astronomical Research in the Southern Hemisphere, Karl-Schwarzschild-Stra{\ss}e 2,  85748 Garching, Germany
  \and Department of Astronomy, University of Minnesota, Minneapolis, MN 55455, USA
  \and School of Physics and Astronomy, The University of Leeds, Leeds, LS2 9JT, UK} 

\abstract {}{The role of episodic mass loss is one of the outstanding questions in massive star evolution. The structural inhomogeneities and kinematics of their nebulae are tracers of their mass-loss history. We conduct a three-dimensional morpho-kinematic analysis of the ejecta of $\eta$~Car outside its famous Homunculus nebula.}{We carried out the first large-scale integral field unit observations of $\eta$~Car in the optical, covering a field of view of 1\arcmin$\times$1\arcmin\ centered on the star.  Observations with the Multi Unit Spectroscopic Explorer (MUSE) at the Very Large Telescope (VLT) reveal the detailed three-dimensional structure of $\eta$~Car's outer ejecta. Morpho-kinematic modeling of these ejecta is conducted with the code SHAPE.}{The largest coherent structure in $\eta$~Car's outer ejecta can be described as a bent cylinder with roughly the same symmetry axis as the Homunculus nebula. This large outer shell is interacting with the surrounding medium, creating soft X-ray emission. Doppler velocities of up to $3\,000$~km~s$^{-1}$ are observed.  We establish the shape and extent of the ghost shell in front of the southern Homunculus lobe and confirm that the NN condensation can best be modeled as a bowshock in the orbital/equatorial plane.}{The SHAPE modeling of the MUSE observations provides a significant gain in the study of the three-dimensional structure of $\eta$~Car's outer ejecta. Our SHAPE modeling indicates that the kinematics of the outer ejecta measured with MUSE can be described by a spatially coherent structure, and this structure also correlates with the extended soft X-ray emission associated with the outer debris field. The ghost shell just outside the southern Homunculus lobe hints at a sequence of eruptions within the time frame of the Great Eruption from 1837--1858 or possibly a later shock/reverse shock velocity separation. Our 3D morpho-kinematic modeling and the MUSE observations constitute an invaluable dataset to be confronted with future radiation-hydrodynamics simulations. Such a comparison may shed light on the yet elusive physical mechanism responsible for $\eta$~Car-like eruptions.}


\maketitle

\section{Introduction}  
\label{intro}

Very massive stars with initial masses above $100~M_{\odot}$ may lose much of their mass in violent pre-supernova (SN) eruptions, which profoundly affect the appearance of spectra and lightcurves in the subsequent SN \citep{2014ARA&A..52..487S}.
In several cases giant outbursts during the Luminous Blue Variable (LBV) phase have been confused with low-luminosity SNe~IIn \citep{2012ASSL..384..249V}. 
Eta~Car's 1843 ``Great Eruption'' is the most famous of such events and the star has become the proto-type of the so-called SN impostors (\citealt{2005ASPC..332...47V,2012ASSL..384.....D}).
In a few cases, LBV-like outbursts have been followed by a genuine core-collapse SN \citep{2007ApJ...666.1116S,2007Natur.447..829P,2012ApJ...744...10K,2014ApJ...789..104O,2014MNRAS.439.2917M}. 

Eta~Car is a binary, and a very luminous and massive system (see \citealt{2012ASSL..384.....D} for a recent summary, and references therein). The star is surrounded by a complex circumstellar environment, which is among the angularly largest of all LBVs. Its famous bipolar nebula, called the Homunculus, was ejected in the 1840s during the Great Eruption. It has an extent of about 18\arcsec\ (0.2~pc) and contains a mass of 10--35~$M_{\odot}$ \citep{2007ApJ...655..911S}. The Homunculus expands with velocities of up to 650~km~s$^{-1}$. Several models for its formation exist, which include stellar interior and atmospheric instabilities, a fast rotating star, and binary interaction (e.g., \citealt{1999ApJ...520L..49L,1998AJ....116..829D,1995ApJ...441L..77F,2010MNRAS.402.1141G,2001A&A...372L...9M,2007ApJ...666..967S,2010ApJ...723..602K}). A smaller bipolar nebula, called the Little Homunculus, was likely ejected in the 1890s and has about 1~$M_{\odot}$ \citep{2003AJ....125.3222I,2004ApJ...605..405S}. It is contained within the larger Homunculus.

\begin{figure}
\centering 
\resizebox{\hsize}{!}{\includegraphics{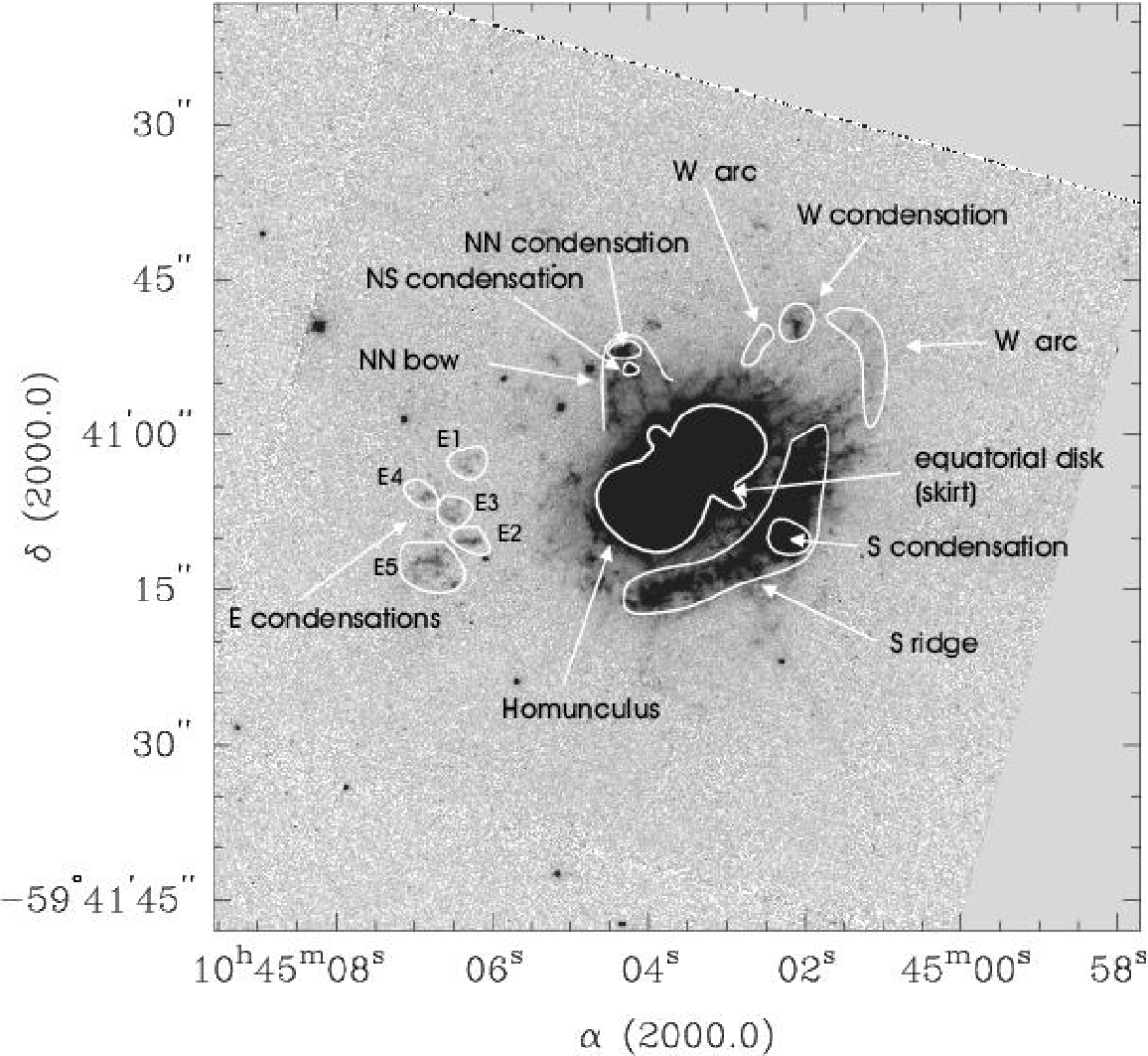}}
\caption{{\it HST/ACS\/} F658N filter image, reproduced from \citet{2004A&A...415..595W}. The field of view is 1.5\arcmin$\times$1.5\arcmin. Several distinct features are identified with the terminology used by \citet{1976ApJ...204L..17W}. The main filter transmission is between $\lambda$6549--6623\AA, ranging over 3\,000~km~s$^{-1}$ in velocity space. Eta~Car's outer ejecta have a large range of velocities and thus [\ion{N}{II}] $\lambda\lambda$6548,6583 and H$\alpha$ emission contribute to the features observed here.}
\label{fig:nomenclation}
\end{figure}

Beyond the bipolar Homunculus there are additional nebulous features referred to as the {\it outer ejecta} with a total mass of $>$2--4~$M_{\odot}$ (see \citealt{2012ASSL..384..171W} for a review and Figure \ref{fig:nomenclation}). The outer ejecta are distributed in a region of 60\arcsec\ (0.7~pc) in diameter and include a variety of structures of different sizes and morphologies \citep{1976ApJ...204L..17W,1999A&A...349..467W}. 
A number of puzzling structures exist, e.g., highly collimated filaments and a jet-like feature. 
A global bipolar expansion pattern also of these outer ejecta was recognized, but the morphology appears irregular.
Kinematic analysis showed the bi-directional expansion of the outer ejecta with velocities between 400--900~km~s$^{-1}$ and the same symmetry axis as the Homunculus \citep{2001A&A...367..566W}. 
An expansion shell in front of the southeast lobe of the Homunculus also matches the global expansion pattern \citep{2002A&A...389L..65C}. 
 
The 19th century Great Eruption of $\eta$~Car ejected more than $10~M_{\odot}$ at speeds of 650~km~s$^{-1}$, with a kinetic energy of about $10^{50}$~erg \citep{2003AJ....125.1458S}, compared to $10^{50}$--$10^{51}$~erg in kinetic energy for a typical core-collapse SN.
Some features in the outer ejecta are moving at up to 3\,500--6\,000~km~s$^{-1}$ \citep{2008Natur.455..201S}, which roughly doubles the total kinetic energy and suggests that the eruption released a blast wave, which is led by a shock front of compressed gas.  The high velocities cause strong interaction with the surrounding medium and give rise to soft X-ray emission. As a result, $\eta$~Car's outer shell mimics a low-energy SN remnant.
The eruption may have been powered by a deep-seated explosion rivaling a SN, perhaps triggered by pulsational pair instability \citep{2002ApJ...567..532H}. The pulsational pair instability induces strong pulsations of the entire star and part of the outer envelope is ejected \citep{2002ApJ...567..532H,2007Natur.450..390W,2008ApJ...685.1103W}. The number of pulsations, the time scales between them (several weeks to a few years), and the mass ejected, depend on the mass of the CO core \citep{2016MNRAS.457..351Y}. Eta~Car's brief brightening events during the Great Eruption may be associated with such pulsations (see \citealt{2011MNRAS.415.2009S} for the historical lightcurve).  Most models for $\eta$~Car's Great Eruption, however, assume a radiation-driven super-Eddington wind \citep{1994PASP..106.1025H,1997ARA&A..35....1D,2000ApJ...532L.137S,2004ApJ...616..525O,2016JPhCS.728b2008D}.

While the Homunculus is mainly a reflection nebula, the outer ejecta are emission nebulae. Chemical composition analysis revealed an overabundance of nitrogen from CNO processed material in the outer ejecta \citep{1982ApJ...254L..47D,1986ApJ...305..867D}, supporting their formation during the evolved stellar phase. However, if the star undergoes tidal mixing in a binary or is a fast rotator, apparent overabundances may be due to enhanced mixing. \citet{2004ApJ...605..854S} find that the more distant clumps have a lower nitrogen abundance. This could be explained by interactions of younger ejecta with unprocessed material from the previous stellar mass loss, but the outer ejecta might also be the relic of an earlier eruptive phase \citep{2004ApJ...616..976G}. A comprehensive model for the formation of the outer ejecta is still missing.

We investigate the outer ejecta of $\eta$~Car using data obtained with the Very Large Telescope Multi Unit Spectroscopic Explorer (VLT MUSE; \citealt{2010SPIE.7735E..08B}). This work is complementary to recent work on the three-dimensional structure of the Homunculus \citep{2014MNRAS.442.3316S}.
\citet{1976ApJ...204L..17W} was the first to review the detailed structure of $\eta$~Car's outer ejecta and his terminology is still in use today. Figure \ref{fig:nomenclation} provides a map to some of the features discussed in this paper. This narrow-band image does, however, not reveal the detailed morphology of the outer ejecta discussed in this paper, because of their large velocity range and line blending.

In Section \ref{data} we describe the MUSE observations and the modeling with the three-dimensional morpho-kinematic code SHAPE  \citep{2011ITVCG..17..454S}. In Section \ref{results} we present our results on the structures of the outer ejecta, and explore the spatial correlation between them and the extended X-ray emission around $\eta$~Car. In Sections \ref{discussion} and \ref{conclusion} we discuss our findings and present our conclusions.

\section{Data and analysis} 
\label{data}

\subsection{Observations}
Eta~Car was observed with VLT MUSE on three nights (2014 November 13, 2014 December 12, and 2015 January 8).
MUSE is an Integral Field Spectrograph, composed of 24 integral field units (IFUs) that sample a continuous 1\arcmin$\times$1\arcmin\ field of view. The instrument covers most of the optical and part of the near-infrared domain (4\,800--9\,300~\AA) with a spectral resolving power of $R\sim1\,600$ (blue) to $R\sim3\,600$ (red), corresponding to a median velocity resolution of $\sim$100~km~s$^{-1}$. The spatial sampling for MUSE's currently offered wide field mode is 0\farcs2. We obtained seeing-limited spatial resolutions between 0\farcs8 and 1\farcs3. 

Eta Car's circumstellar material with an extent of about 60\arcsec, velocities of several hundred to a few thousand km~s$^{-1}$, and numerous optical emission lines is ideally suited for observations with MUSE. However, the bright central source causes strong artifacts (vertical and horizontal stripes, concentric rings) at all wavelengths, probably due to internal reflections in the spectrographs and the fore-optics module. These artifacts are located close to the central star and span mostly over the extent of the Homunculus. They are spatially stable (i.e., they occur in all velocity channels at the same position and, therefore, cannot be confused with intrinsic features of the nebulosities), but vary in intensity across the wavelength range of the reduced single MUSE data cube, where the intensity variations are driven by the spectrum of the bright offending source.
We remove these artifacts from the different emission line channel maps presented here by subtracting median-averaged white light images constructed from nearby emission line free regions.
The rapid intensity variations (with wavelength) of the artifacts renders this procedure difficult and some residual artifacts remain.
The data also show the usual diffraction spikes around bright sources.
Our treatment of the data leaves residuals. However, because these are spatially stable, they can be easily identified and avoided in our analysis.

\begin{figure*}[p]
\centering
  \includegraphics[width=1\linewidth]{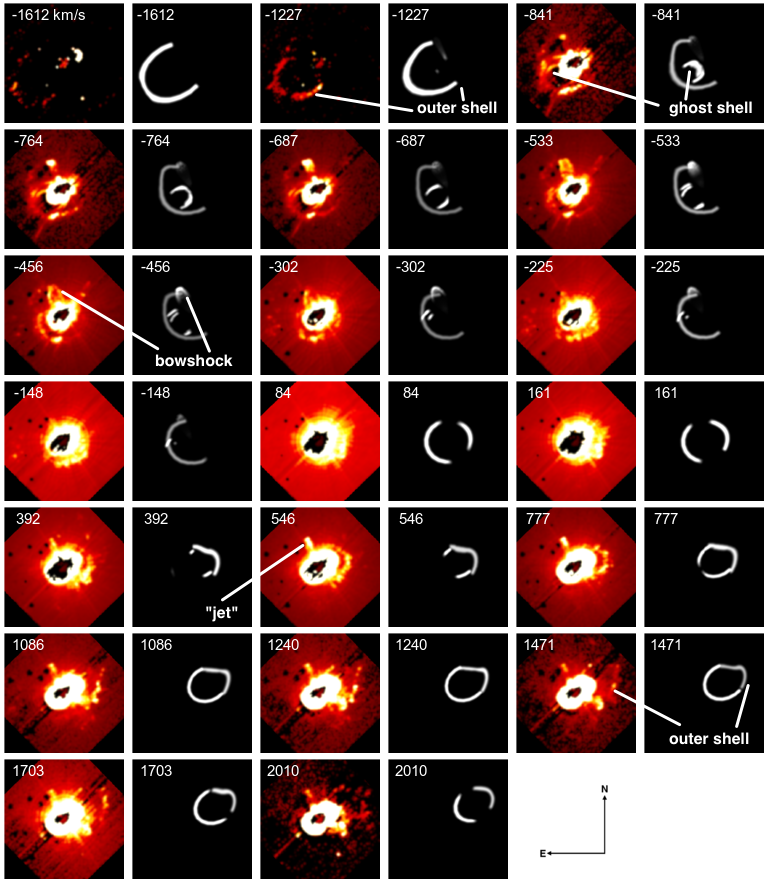}
  \caption{MUSE channel maps showing the observed H$\beta$ emission in the MUSE field of view of 1\arcmin$\times$1\arcmin. Each panel shows a single channel of the MUSE data cube, corresponding to different velocities of the H$\beta$ emission. The velocities are indicated in each panel in units of km~s$^{-1}$ (each channel map spans a range of about 80~km~s$^{-1}$, velocities are not corrected for $\eta$~Car's systemic  heliocentric velocity of roughly $-$8~km~s$^{-1}$). The corresponding SHAPE model channel map is shown on the right side of each observational channel map (the grayscale only indicates the location of the emission and does not reproduce the brightness distribution), see text for details. The channel maps are irregularly spaced in velocity to show best the brightest emission features corresponding to the outer shell, the ghost shell, the bowshock, and the ``jet'', identified by labels and arrows pointing toward the feature (see also Figure \ref{fig:largescale} for their three-dimensional location). The ``jet'' is not modeled with SHAPE, since the origin of the emission is not clear, see Section \ref{jet}. The central star and part of the Homunculus are saturated. The channel maps are presented with a square-root color stretch and normalized in flux individually to peak intensity. They only contain physical information on the structure and velocities, but do not permit channel-to-channel comparisons to be made of the fluxes.
}
\label{fig:channelmaps}
\end{figure*}

MUSE provides the first complete and velocity resolved set of optical line ratios and line centroids at each position of the large-scale nebula around $\eta$~Car.
Observations with exposure times from 0.02~s to 300~s were obtained with $\eta$~Car in the center of the field of view. Additional exposures of 300~s and 600~s were obtained at offset positions 35\arcsec\ and 70\arcsec\ along the Homunculus axis and perpendicular to it. In this paper, we only use the observations centered on $\eta$~Car, covering a field of view of $1\arcmin \times 1\arcmin$. The channel maps shown in Figure \ref{fig:channelmaps} are extracted from a 300~s exposure. 

The data were reduced using version 1.1.0 of the MUSE ESO standard pipeline \citep{2012SPIE.8451E..0BW}.
Bias, arc, and flatfield master calibration files were created using the default set of calibrations exposures provided by ESO. These include an illumination correction flatfield taken within one hour of the science observations to correct for potential temperature variation in the illumination pattern of the slices.\footnote{The MUSE field of view is split into 24 slices, which are sent to separate IFUs. These 24 channels are further sliced into 48 $15\arcsec \times 0.2\arcsec$ slitlets.}
All data cubes were resampled to $0\farcs2\times0\farcs2\times1.25~\AA$. 

QFitsView\footnote{\url{http://www.mpe.mpg.de/~ott/dpuser/qfitsview.html}} was used to extract channel maps, position-velocity diagrams, and circular aperture spectra. Figure \ref{fig:channelmaps} shows selected channel maps of the H$\beta$ emission in $\eta$~Car's outer ejecta. Each panel displays a single wavelength channel of the MUSE data cube, corresponding to a narrow velocity range of about 80~km~s$^{-1}$. We analyzed for a few selected lines the velocity space from $-6\,000$ to $6\,000$~km~s$^{-1}$. 
H$\alpha$ and the nearby [\ion{N}{II}] lines at $\lambda\lambda$6548,6583 are very strong in emission in $\eta$~Car's outer ejecta and can have similar line peak strengths. 
Line blending due to the large radial velocities of the ejecta and multiple components per line of sight is a serious issue for the [\ion{N}{II}] and H$\alpha$ lines. [\ion{N}{II}] $\lambda$6548 is $-675$~km~s$^{-1}$ and [\ion{N}{II}] $\lambda$6583 is $945$~km~s$^{-1}$ in velocity space from H$\alpha$, very common velocities in the outer ejecta. Clearly disentangling each of these lines is a challenging task.  
Here, we thus chose the H$\beta$ emission line as the primary line to analyze the three-dimensional morpho-kinematic behavior of $\eta$~Car's outer ejecta. Another suitable, but fainter, emission line for this type of work is [\ion{S}{III}] $\lambda$9069. Both lines have only little contamination from nearby lines of \ion{Fe}{II}, [\ion{Fe}{II}], \ion{Cr}{II}, and \ion{N}{I}. The channel maps shown in Figure \ref{fig:channelmaps} are presented with a square-root color stretch and were normalized in flux individually to highlight the location of the outer (and fainter) ejecta. Absolute brightness distribution is not required for the work presented here, which focuses on the structure and dynamics of the outer ejecta.
  
Unless mentioned otherwise, the velocities in the following sections are radial velocities and the distances are the values projected onto the plane of the sky. Velocities are not corrected for $\eta$~Car's systemic velocity of roughly $-$8~km~s$^{-1}$ (heliocentric; \citealt{1997AJ....113..335D,2004MNRAS.351L..15S}). Throughout the paper we assume a distance to $\eta$~Car of 2.35~kpc \citep{1993PASAu..10..338A,2001AJ....121.1569D,2006ApJ...644.1151S}. Quoted wavelengths are air values.

\subsection{Morpho-kinematic modeling with SHAPE}

\begin{figure*}
\centering
  \includegraphics[width=1\linewidth]{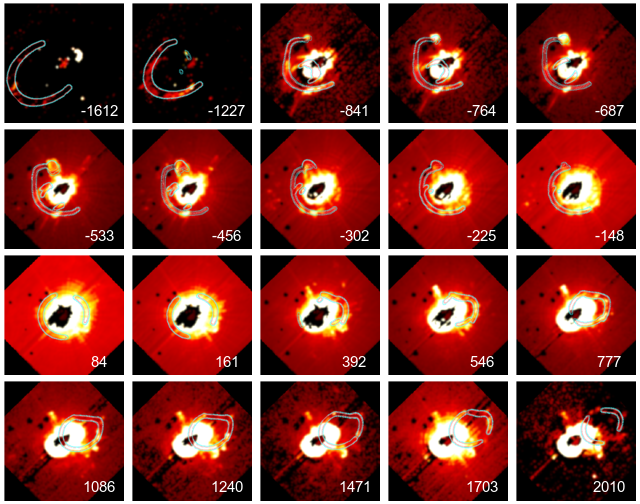}
  \caption{MUSE channel maps showing the observed H$\beta$ emission in the MUSE field of view of 1\arcmin$\times$1\arcmin\ as in Figure \ref{fig:channelmaps}, but overlayed with the brightness outlines of the model features. Only a few contours are shown to not obstruct the underlying observation.}
\label{fig:channelmaps+modelcontours}
\end{figure*}

From the spatio-kinematic MUSE data, we construct a three-dimensional geometry of $\eta$~Car's outer ejecta with the morpho-kinematic modeling code SHAPE \citep{2011ITVCG..17..454S}. SHAPE has been used extensively to model the complex three-dimensional structure of planetary nebulae. Recently,  \citet{2014MNRAS.442.3316S} constructed the first three-dimensional model of the Homunculus based on kinematic molecular hydrogen data from VLT X-shooter data. The X-shooter data cover the entire Homunculus nebula, but none of the outer ejecta.

SHAPE enables the user to interactively optimize model parameters by direct comparison of model predictions with observations. The general modeling procedure is described in \citet{2011ITVCG..17..454S} and on the SHAPE support website.\footnote{\url{http://www.astrosen.unam.mx/shape/}}
The program consists of a three-dimensional modeling view in which the geometry and behavior of the model are defined, and a two-dimensional view where the simulated appearance of the model can be compared to observational data.
SHAPE models are constructed interactively with three-dimensional structural mesh elements, such as spheres or cylinders, based on the user's assumptions about the object's geometry, and the emission and the velocity information extracted from observations. Small-scale structure can be added in a variety of ways by modifying the elementary meshes. The mesh models can also be combined with hydrodynamic simulations performed with the corresponding SHAPE module. We have used this technique to model the bowshock structure to the northeast of the Homunculus. 
The three-dimensional model is the input to a rendering module, which emulates how the object would be observed with an imager or spectrograph. 
The user can then interactively refine the model until it qualitatively fits the observational data. In the case of our MUSE data, we modeled several large-scale features, which were selected based on their apparent spatial association as a contiguous structure in the observational data. This includes the larger shells, the ghost shell, and the bow-shock-like feature, described in Section \ref{results}. Other, small-scale features were not included, because they do not provide additional constraints for our morpho-kinematic modeling.
Quantitative fits of SHAPE models to our MUSE observations are not possible within the SHAPE tool. However, such an analysis is not required here, because we are limited by other factors described in the following paragraph.
Given an assumption for the velocity field, the model solution is unique with an additional geometric constraint, such as, e.g., an approximately circular cross-section. 

A cautionary note: Reconstruction based on Doppler-velocity measurements of expanding nebulae often assumes homologous expansion, i.e., radial expansion with the velocity increasing linearly with distance ($\vv{v}  \propto \vv{r}$). 
We adopt homologous expansion for the structures analyzed in this paper with the exception of the bowshock-like feature discussed in Section \ref{jet}, where a hydrodynamical simulation was performed. 
This is because the events that produced the structures were short compared to the dynamic time scale of the expansion, thereby resembling more a ballistic blast wave than a continuous wind shock. A homologous flow may not be valid in detail or throughout the expanse. Ejecta may be slowed down when hitting older material and by interactions with the environment or may have been accelerated by a blast wave. Interaction with preexisting density fluctuations alter the velocity field and distort the reconstructions to an extent that depends on the time-scale of the interaction \citep{2009ApJ...691..696S}. Assuming an incorrect velocity field will lead to distortions in the reconstructed shape. Refinements of the three-dimensional structure model will require improvements of the constraints on the ejection time and velocity field.  {\it Our reconstruction of the outer ejecta thus only gives a suitable first order approximation.}
We urge the reader to keep this in mind.

\section{Results}
\label{results}

\begin{figure*}[!]
\centering
\vspace*{0.5cm}
\begin{minipage}{.5\textwidth}
  \centering
\textbf{\fbox{Earth View}}
  \includegraphics[width=1\linewidth]{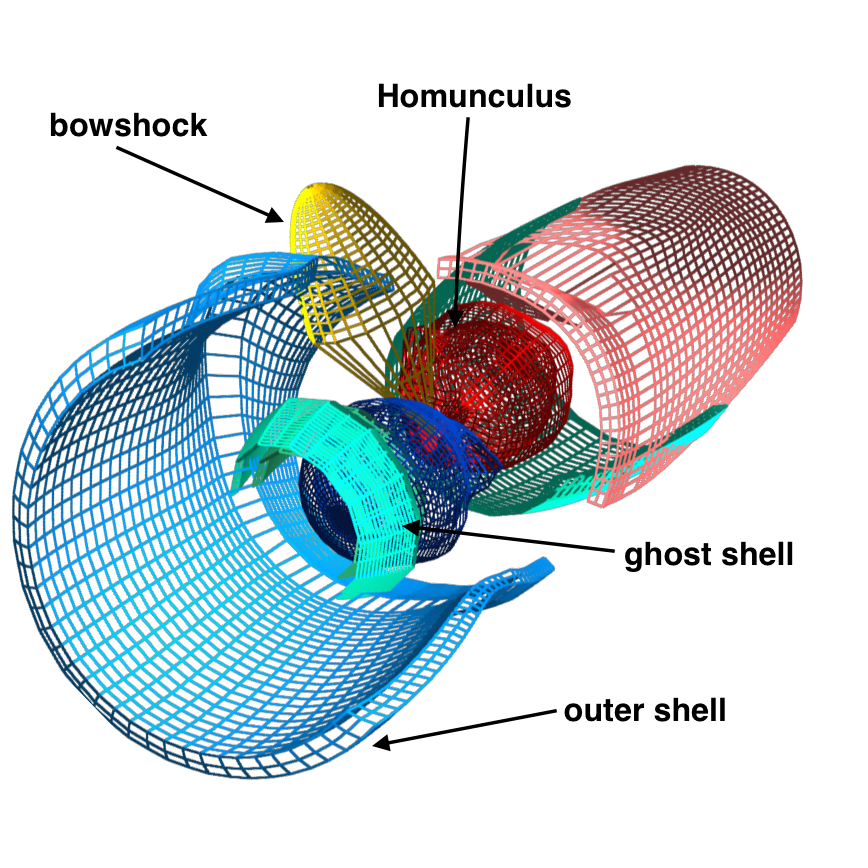}
\end{minipage}%
\begin{minipage}{.5\textwidth}
\vspace*{-2.5cm}
\textbf{\fbox{Side View}}
  \centering
  \includegraphics[width=1\linewidth]{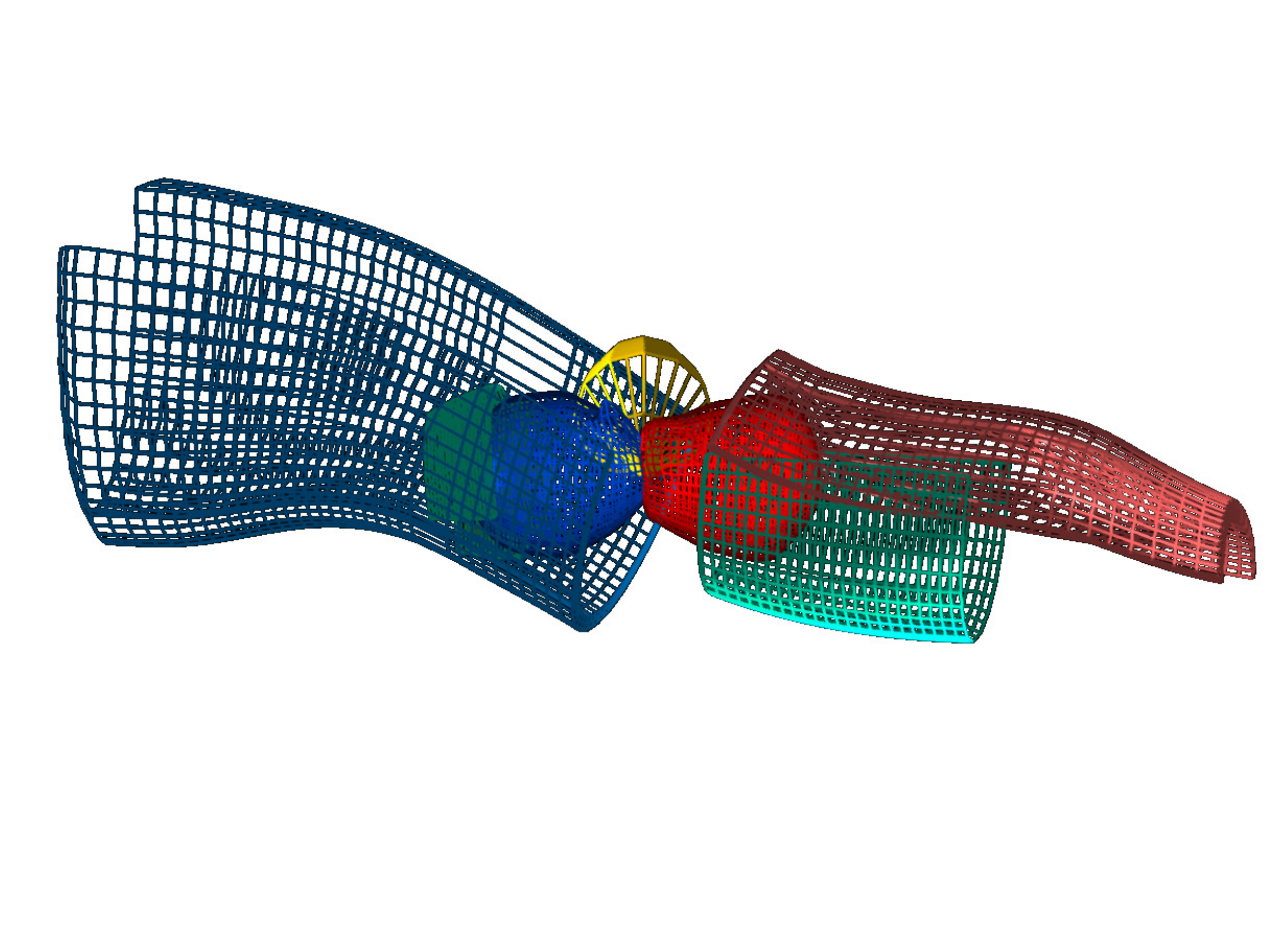}
\end{minipage}
  \caption{Three-dimensional model of $\eta$~Car's outer ejecta seen from Earth (left) and from the side (right) with the bipolar Homunculus nebula in the center (dark red and blue lobes). The model has been constructed with SHAPE and the H$\beta$ emission and velocity (over a range of $\pm3\,000$~km~s~$^{-1}$) information extracted from the MUSE data. The outer ejecta form a partial bent cylinder. The open cylinder parts are not observed in the MUSE data, i.e., we only trace the bottom of the shell moving toward us and the top of the shell moving away from us. The open cylinder parts are likely too faint, maybe due to only weak interactions with less dense surrounding material. This large tube is probably bent or indented due to material being slowed down by interactions with the environment. The ghost shell (green) is in our line of sight to the southern Homunculus lobe and consists of two closely aligned shells. An additional half-shell (green mesh, complementing the bottom of the large cylindrical shell moving away from us) lies behind the northern Homunculus lobe. The bowshock-like feature is displayed in yellow. The colors are selected to distinguish the different shapes selected for the SHAPE modeling.}
\label{fig:largescale}
\end{figure*}

\begin{figure*}[!]
\centering
\vspace*{0cm}
\begin{minipage}{.5\textwidth}
  \centering
  \includegraphics[width=1\linewidth]{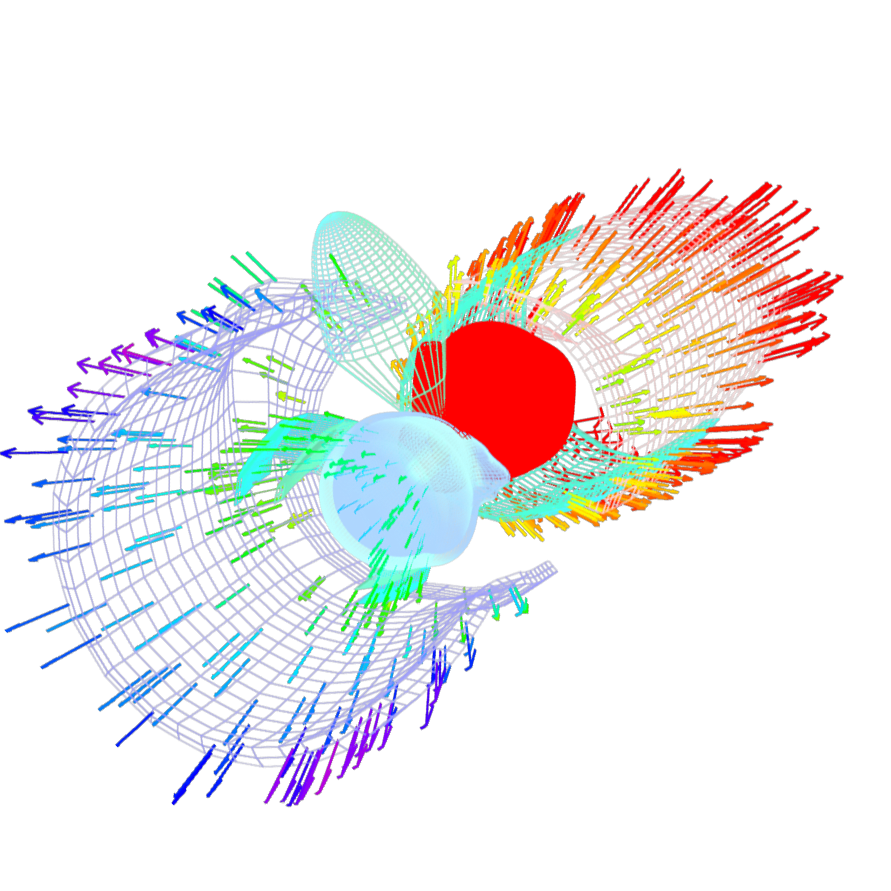}
\end{minipage}%
\begin{minipage}{.5\textwidth}
  \centering
  \includegraphics[width=1\linewidth]{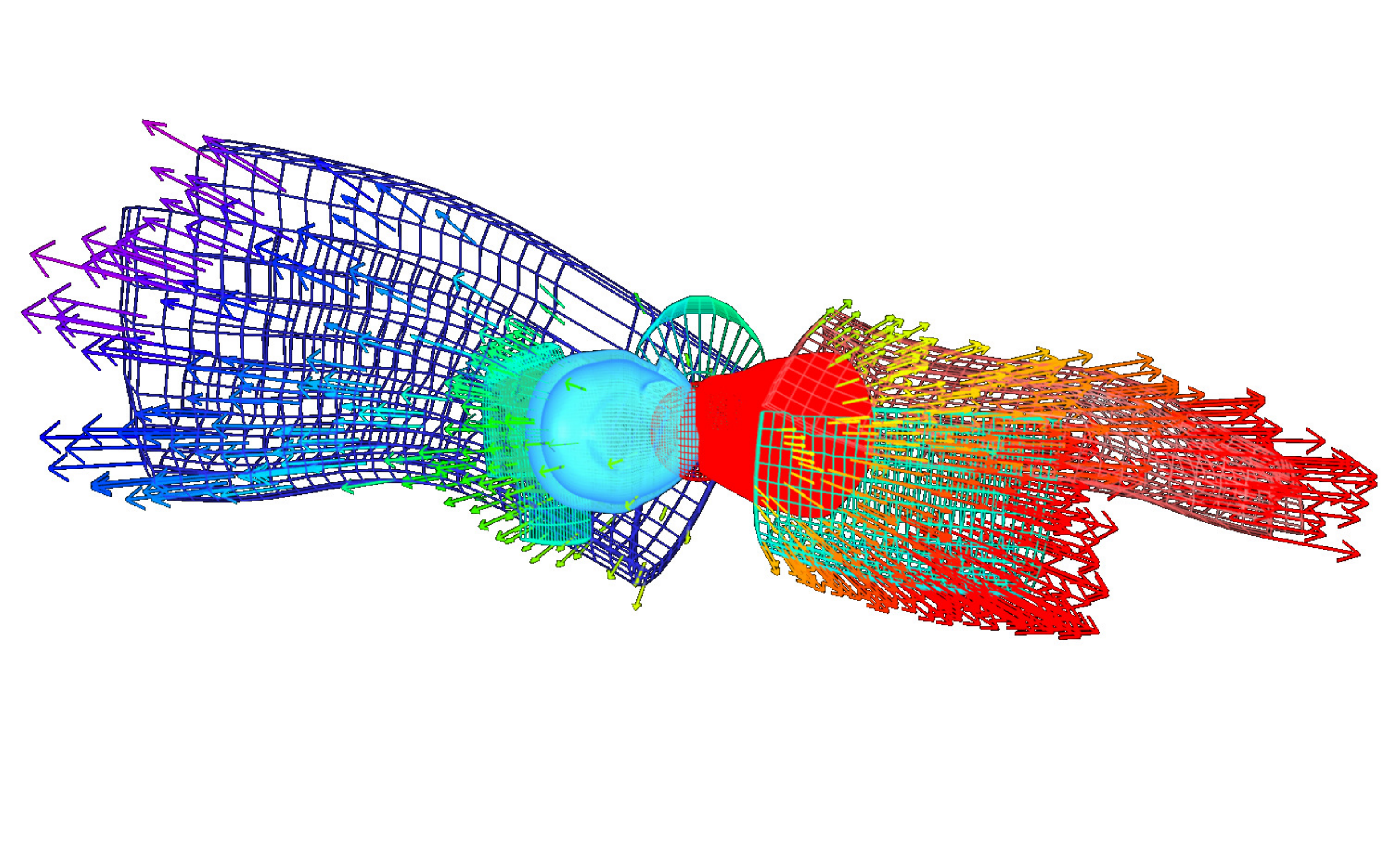}
\end{minipage}
  \caption{Three-dimensional velocity structure of $\eta$~Car's outer ejecta seen from Earth (left) and from the side (right). The color coding is from $-$2\,500~km~s$^{-1}$ (blue/violet) to +2\,500~km~s$^{-1}$ (red). This SHAPE model shows only qualitatively where material is located. The colors of the velocity vectors and the vector lengths show roughly the direction and speed of the material.}
\label{fig:velscale}
\end{figure*}

With MUSE IFU observations we uncover the apparently irregular structure  of $\eta$~Car's outer ejecta and show for the first time their spatially contiguous large-scale morphology. 
In the past, narrow-band images excluded the high velocity ejecta and contained emission from more than one atomic transition. This made the detection of a contiguous morphology difficult. Broad-band images were dominated by the bright central source and the Homunculus nebula, and were not deep enough to show the structure of the outer ejecta. Unlike the bipolar Homunculus, $\eta$~Car's outer ejecta thus did not appear symmetric or connected (Figure \ref{fig:nomenclation}), but seemed to be composed of many irregularly shaped and randomly placed structures in an elliptically shaped region (described as filaments, jets, arcs, bullets or knots, and strings). Integrating the MUSE data over the {\it HST\/} WFPC2 filter transmission curves we identify the same features observed in the {\it HST\/} images  (filtered to match the angular resolution of MUSE) and can obtain their proper motions with a baseline of 17.5~yr by comparing 1997 {\it HST\/} images with our 2014 MUSE data. We use proper motion measurements to constrain the ejection time of the NN bow, one of the most prominent features in $\eta$~Car's outer ejecta and discussed in Section \ref{jet}.

MUSE provides the first complete spatio-spectral sampling of $\eta$~Car's outer ejecta. The outer ejecta are indeed more clumpy than the material in the Homunculus, but most of the knots outline a coherent superstructure.
Figures \ref{fig:channelmaps} and \ref{fig:channelmaps+modelcontours} show selected observed MUSE velocity channel maps and the corresponding SHAPE model channel maps.
Rough morpho-kinematic modeling with SHAPE of the emission features visible in the MUSE position-velocity diagrams across the H$\beta$ emission line reveal a continuous morphology of $\eta$~Car's outer ejecta that is not a simple bipolar shell.  The three-dimensional model is shown in Figures \ref{fig:largescale} and \ref{fig:velscale} and described in detail in the next subsections. The largest continuous structure surrounding $\eta$~Car can be identified as the ``cocoon'' or ``blast wave'' described in \citet{2004ApJ...605..854S} and \citet{2008Natur.455..201S}. We also determine approximately the shape and extent of the so-called ghost shell in front of the southeast Homunculus lobe \citep{2002A&A...389L..65C}. In addition, we discuss the jet-like NN bow and condensation in the equatorial plane \citep{1993A&A...276L..21M}.

\subsection{Prominent features in the outer ejecta}
\label{description}

Before describing the main large-scale features in $\eta$~Car's outer ejecta (i.e., the outer shell, the ghost shell, and the NN bow) in the following subsections, we give a short description of the prominent features in Figure \ref{fig:nomenclation} and how the MUSE data improves their interpretation. We follow the terminology used by \citet{1976ApJ...204L..17W}.

\begin{itemize}

\item The {\it E condensations} are five discrete knots southeast of the Homunculus \citep{1976ApJ...204L..17W,2012ASSL..384..171W}. The MUSE data reveal the velocity structure of these condensations. They have a range of negative velocities. The condensation E5 has the fastest velocities, with speeds of $v \sim -$600~km~s$^{-1}$. It consists of several knots and arc-like features. E2 and E4 can be observed up to velocities of $v \sim -$450~km~s$^{-1}$. E2 and E3 can be better described as arcs than knots and are visible up to velocities of $v \sim -$300~km~s$^{-1}$. They can be seen in several of the channel maps in Figure \ref{fig:channelmaps}. The E condensations are not part of the continuous shell discussed in Section \ref{shape}, such that their nature remains particularly mysterious.

\item In narrow-band images, the {\it S ridge} appears to be the largest structure in $\eta$~Car's outer ejecta \citep{1976ApJ...204L..17W,2012ASSL..384..171W}. The S ridge is composed of many filaments and knots  with velocities between $v \sim -$300~km~s$^{-1}$ in the southwest and up to more than  $v \sim 1000$~km~s$^{-1}$ in the northwest.  MUSE data show that this ridge is part of the large outer shell discussed in Section \ref{shape}. The S ridge and S condensation have not been modeled as separate features, because they appear to be part of the overall structure of the outer shell (Figures \ref{fig:channelmaps} and \ref{fig:largescale}).

\item The {\it W arc}  \citep{1976ApJ...204L..17W,2012ASSL..384..171W} is also part of the same outer shell structure and can be observed up to velocities of $v \sim $~2000~km~s$^{-1}$. The {\it W condensation}, on the other hand, is a distinct knot at smaller radial velocities around $v \sim $~350~km~s$^{-1}$.

\item The {\it NN, NS condensations}, and the {\it NN bow} are intriguing features, which will be discussed in detail in Section \ref{jet}. They lie within the equatorial plane and can be observed over a wide range of negative and positive velocities ($\pm1000$~km~s$^{-1}$).

\item Five long, highly collimated linear features, called {\it strings} \citep{1999A&A...349..467W}, are perhaps some of the most interesting structures in the outer ejecta.
They are found in the northwest and southeast, pointing radially away from the Homunculus (they are faint and thus not visible with the cuts used in Figure \ref{fig:channelmaps}). Their velocity structure is resolved in the MUSE data. The velocities increase with distance up to $v \sim -$1000~km~s$^{-1}$. The MUSE data do not allow to constrain their three-dimensional motion better than previous work \citep{1999A&A...349..467W}. 

\end{itemize}

\subsection{The outer shell}
\label{shape}

MUSE channel maps across H$\beta$ reveal a large shell-like emission feature with  velocities between $-2\,200$~km~s$^{-1}$  southeast of the central source and $+2\,000$~km~s$^{-1}$  northwest of the central source (Figure \ref{fig:channelmaps}). At a velocity of $-2\,200$~km~s$^{-1}$, the shell has an extent of about 30\arcsec\ toward the southeast of the central star. For velocities between $-2\,200$~km~s$^{-1}$ and $0$~km~s$^{-1}$, the emission of this shell appears to gradually move toward the central source. The shell is oriented along a similar direction as the southern Homunculus lobe. At positive velocities, a complementary shell structure becomes evident northwest of the central star, which can be traced up to a distance of about 30\arcsec\ and velocities up to $+2\,000$~km~s$^{-1}$. With deeper images, faster velocities may be detected and the structure may be found to extend beyond what is visible in our MUSE data set.

In the MUSE data we measure the radial velocities of the ejecta and the projected distances. We then have to adopt an ejection time in order to determine the distances of material from the star -- unless we assume a priori a geometry. For the SHAPE modeling of this large shell-like structure seen in emission, we use a homologous expansion (i.e., material is neither decelerated nor accelerated) for the ejecta and set the ejection date at the time of the Great Eruption.\footnote{The selected velocity field corresponds to an ejection of the material in 1836. Within the accuracy of the model and observations, an ejection date of 1843 could be used as well.} 
We adopt the same age for the outer ejecta as for the Homunculus (i.e., ejected in the same physical event) for the following reason. 
If we assume a homologous flow with the geometrical constraint that the tube of the outer ejecta has roughly a circular cross-section, we find an age for the outer ejecta 1.4 times older than the Homunculus. But no major eruptions were reported in $\eta$~Car in the two centuries before the Great Eruption (\citealt{2011MNRAS.415.2009S}, though only few data points are recorded from the 17th century).

With SHAPE we derived that the emission of the outer shell can be matched with a bent cylinder geometry embedding the entire Homunculus nebula (Figure \ref{fig:largescale}). This structure is reminiscent of the shape of the planetary nebula M~2-9 \citep{2015A&A...582A..60C}. The bending of the outer shell might have been caused by the interaction with a denser region of the ambient medium, slowing down the material. The reconstructed bent might be somewhat exaggerated, because the interaction with surrounding material might have only started recently compared to the age of the outer shell. The actual bending may therefore be less pronounced than depicted in the current  reconstruction, but will increase with time. 
In Appendix \ref{sec:alternative} we show a SHAPE model, which assumes a scaled-up Homunculus geometry for the outer ejecta. It illustrates that a Homunculus geometry does not match the MUSE data.

The emission brightness is irregularly distributed over the extent of the cylinder. Brighter knots are observed and the cylinder has its strongest emission on opposite sides for its blue and red components (see the structure labeled ``outer shell'' in Figure \ref{fig:channelmaps}). The emission originates from the bottom half of the cylinder for the blue component (viewed from our line of sight), while most of the emission comes from the top half for the red component. These bright parts lie along the line where the cylinder wall intersects with the Homunculus axis. That we only observe a partial cylinder is probably because the other parts are too faint to be detected in our data, maybe due to only weak interactions of the ejecta with less dense surrounding material in those directions. The shell depicted as a green mesh behind the northern Homunculus lobe may be completing the tube, but we cannot clearly establish this (Figures \ref{fig:largescale} and \ref{fig:velscale}). 

The orientation of the cylinder spatially close to $\eta$~Car is misaligned with respect to the  Homunculus axis (Figure \ref{fig:largescale}), which has an inclination of 41\degree\ \citep{2001AJ....121.1569D,2006ApJ...644.1151S}. Further away from $\eta$~Car, the inclination of the cylinder is approximately aligned with the Homunculus axis. Projected on the plane of the sky, the cylinder is rotated by approximately $-$10\degree\ with an estimated error of about 3\degree\ with respect to the Homunculus. 
The outer ejecta are not a perfect cylinder, but locally the orientation varies. The difference in orientation with respect to the Homunculus could be a secondary effect of the interaction with an inhomogeneous environment.

The diameter of the cylinder section moving toward us varies between $1.0 \times 10^{18}$~cm and $1.2\times 10^{18}$~cm (0.32--0.39~pc). The diameter of the cylinder section moving away from us is noticeably smaller at about $6.7\times 10^{17}$~cm (0.22~pc). The diameters vary by about 20\% with position. The thickness of the shell is approximately 15\% of its radius.
The material behind the star probably runs into denser material and thus cannot expand as easily as in front of the star, resulting in the smaller cylinder diameter. 
H$\beta$ emission tracing the component in front of the star is smooth and continuous, while the emission of the component behind the star is more irregular and clumpy. There is also more soft X-ray emission in the northwest than in the southeast of $\eta$~Car. This is consistent with  stronger interactions between the shell and the surrounding material on the red side. We will discuss this further in Section \ref{sec:X-rays}.
Note that with SHAPE we can only model the brightness distribution of the ejecta, but do not obtain any information on the density distribution.

\subsection{The ghost shell(s)}

In the MUSE channel maps, two horn-like features can be seen southeast of $\eta$~Car. The first horn-like feature becomes apparent at velocities of about $-900$~km~s$^{-1}$ (labeled ``ghost shell'' in the $-$841~km~s$^{-1}$ observed and model channel maps in Figure \ref{fig:channelmaps}), the second at velocities of about $-600$~km~s$^{-1}$. These emission features can be modeled by two adjoining shells, which lie along our line of sight just in front of the southern Homunculus lobe and are moving toward us (green mesh in Figure \ref{fig:largescale}). Figure \ref{fig:channelmaps} shows a good fit between the MUSE observations and the SHAPE model. 
These two shells appear to align with the southern Homunculus lobe.
Behind the northern Homunculus lobe, we identify a structure that extends about twice as far as the size of the Homunculus with projected velocities larger than $+400$~km~s$^{-1}$ (green mesh in Figure \ref{fig:largescale}). It is not clear if this shell is the counterpart of the ghost shell in front of the southern lobe or part of the outer shell.

We identify the two shells in front of the southern Homunculus lobe with the ghost shell discussed in \citet{2002A&A...389L..65C}. \citet{2002A&A...389L..65C} detected the ghost shell in emission in multiple Balmer lines and in  forbidden lines such as [\ion{N}{II}], [\ion{S}{II}], and [\ion{Ar}{III}] at velocities between $-$675~km~s$^{-1}$ and $-$850~km~s$^{-1}$. It is probably also associated with the complex absorption structure of Ca~H and Ca~K lines \citep{2001AJ....121.1569D} and the $-$513~km~s$^{-1}$ absorption system in the near-ultraviolet \citep{2006ApJS..163..173G}. 
\citet{2002A&A...389L..65C}   hypothesized that the form of the shell is roughly a sphere with radius $\sim$11\arcsec\ that surrounds the Homunculus, but that the quasi-spherical approximation is not valid near the central star and in the northwest lobe region. They propose that the shell is the forward shock between the fast stellar wind of the Great Eruption and the older slow massive stellar wind, distorted by ejecta such as the equatorial disk. 

MUSE enables us to show clearly the structure and extent of the ghost shell (Figure \ref{fig:largescale}). We identify two adjacent shells, which are located just outside the southeast lobe of the Homunculus. They have a similar shape as the southern lobe, but velocities several hundreds of km~s$^{-1}$ faster than the Homunculus material.

\subsection{The NN bow and the ``jet''}
\label{jet}

In the Homunculus equatorial plane, toward the northeast, is a remarkable structure, identified as NN bow and NN condensation (Figure \ref{fig:nomenclation}). \citet{1993A&A...276L..21M} interpreted this prominent knot as the interaction of a jet with the ambient gas.
In the MUSE velocity channel maps, we observe a large feature at negative velocities that is reminiscent of a bowshock, but not the corresponding jet. The emission feature has the structure of a giant loop and has a radial velocity of about $-500$~km~s$^{-1}$ at its furthest extent from $\eta$~Car at the NN condensation. The entire structure has a large velocity range between $-850$~km~s$^{-1}$ and $-200$~km~s$^{-1}$. 
We measured the proper motion of the NN bow between 1997 July and 2014 December using an {\it HST\/} WFPC2 F656N image and our MUSE data integrated over the WFPC2 F656N filter transmission curve. The displacement of the tip of the NN bowshock (the NN condensation) in the 17.5~yr between these two epochs is 1.5--2\arcsec\ (corresponding to a transverse velocity of about 200~km~s$^{-1}$; compare to \citealt{2016MNRAS.463..845K}, who find proper motions of 200~km~s$^{-1}$ to 1400~km~s$^{-1}$ for this region).

Instead of using simple structural elements, we model this bowshock-like feature using the numerical hydrodynamics extension of SHAPE \citep{2013MNRAS.436..470S}. The measured radial velocity and proper motion provide some constraints for the model parameters. We find that the observations can be matched by a dense jet with a spatial bowshock velocity of 1\,300~km~s$^{-1}$ running into an inhomogeneous medium (yellow mesh in Figure \ref{fig:largescale}). The jet originates $10\degree\pm10$\degree\ from the plane of the equatorial skirt toward the blue lobe, roughly perpendicular to the Homunculus axis, and at about $25\degree\pm5$\degree\ from the plane of the sky toward the observer. Its deprojected extent from the star is about $6.9\times 10^{17}$~cm (0.22~pc). Figure \ref{fig:channelmaps} shows the observed channel maps next to our model. We only observe the bowshock, but not the corresponding jet in the MUSE data at negative velocities.

The region where this bowshock-like feature is observed is faint in X-ray emission, which hints at absorption of the soft X-rays by an extension of the equatorial disk or by the bowshock itself \citep{2004A&A...415..595W}. The lack of soft X-rays may also be caused if the material would be expanding into a low-density region. In this case, the bowshock-like structure could be explained by material expanding freely into the low-density ambient medium, in contrast to other outflow directions. 

At the same spatial location, but at positive velocities, a very bright collimated emission resembling a jet is visible (Figure \ref{fig:channelmaps}). Velocities along this ``jet'' increase only slightly with distance to the central source. The emission brightness peaks at 500--600~km~s$^{-1}$. The emission line profile is very broad and extends over about 800~km~s$^{-1}$. A jet at positive velocities cannot cause a bowshock at negative velocities and the relation of this feature to the bowshock-like structure is thus questionable. The spectrum of this ``jet'' shows continuum emission and strong emission from hydrogen and helium lines present in the stellar spectrum, but red-shifted. Thus, this structure results most likely from processes including reflection and/or scattering of light from the central star by an extension of the equatorial skirt. The bright structures observed at negative and positive velocities may not be directly related, but the equatorial plane is a preferred direction for outflows and scattering processes.
This structure has a possible counterpart toward the southwest with similar projected velocities.

\subsection{Correlation of the outer ejecta with the soft X-ray emission}
\label{sec:X-rays}

\begin{figure}
\centering
  \includegraphics[trim=3cm 3cm 3cm 3cm, clip=true, width=0.95\linewidth]{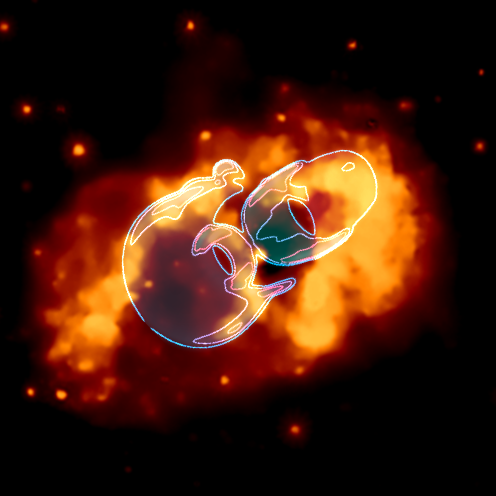}
  \caption{Composite images displaying our SHAPE model together with the 0.5--1.2~keV {\it Chandra\/} X-ray image. The soft X-ray emission is produced as material runs into the nearby gas and dust. X-ray emission from the Homunculus is detected only in hard X-ray bands. The large-scale optical tube structure fits right inside the X-ray bubble. The SHAPE model contours show only qualitative the outlines of the emission. The blue and red colors are to distinguish the different shapes. There is no physical information associated with them. For the image scale, see Figure \ref{fig:xray_vel}.}
\label{fig:xrays}
\end{figure}

\begin{figure*}[!]
\centering
\begin{minipage}{.5\textwidth}
  \centering
  \includegraphics[width=1\linewidth]{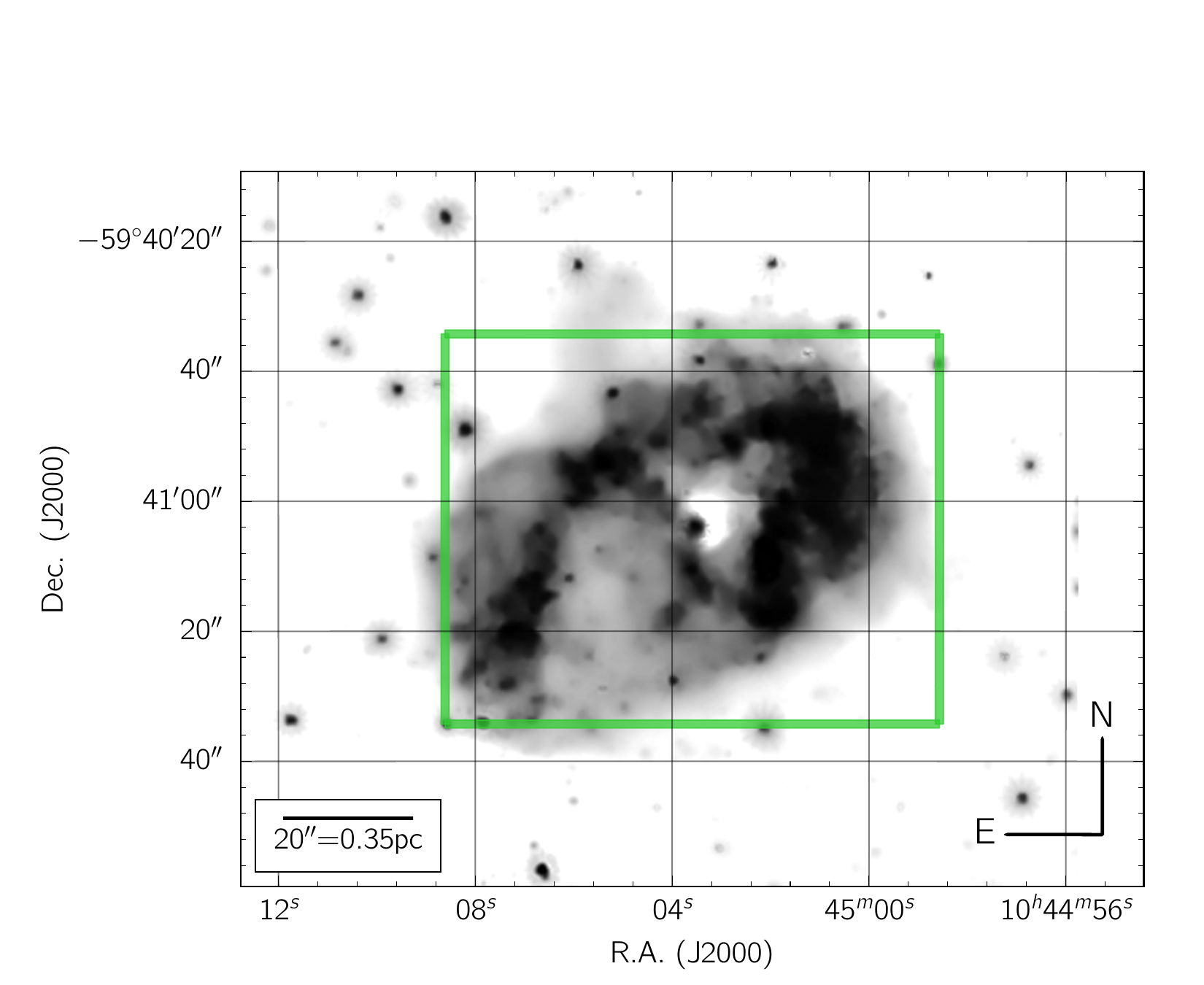}
\end{minipage}%
\begin{minipage}{.5\textwidth}
  \centering
  \includegraphics[width=1\linewidth]{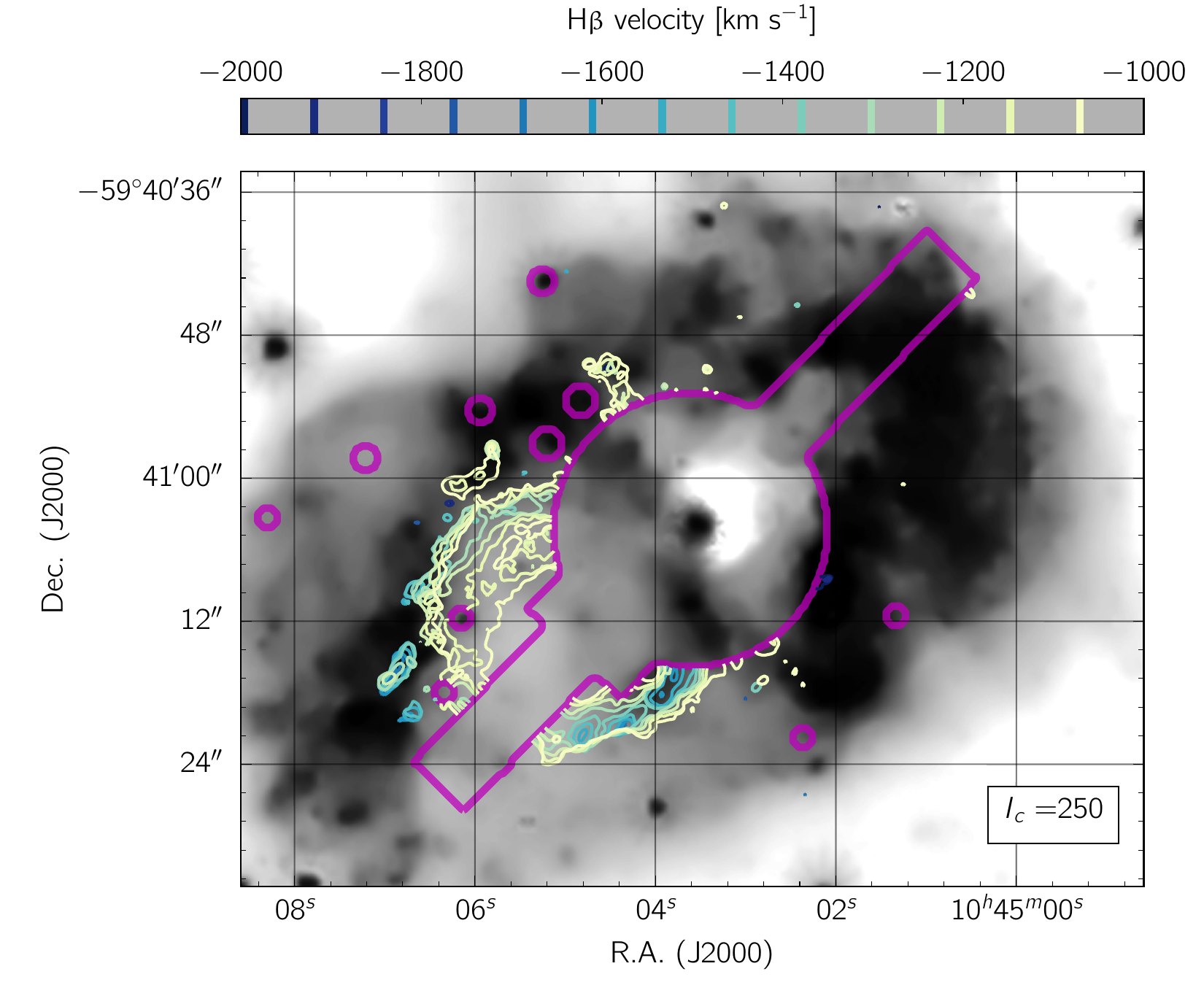}
\end{minipage}
\begin{minipage}{.5\textwidth}
  \centering
  \includegraphics[width=1\linewidth]{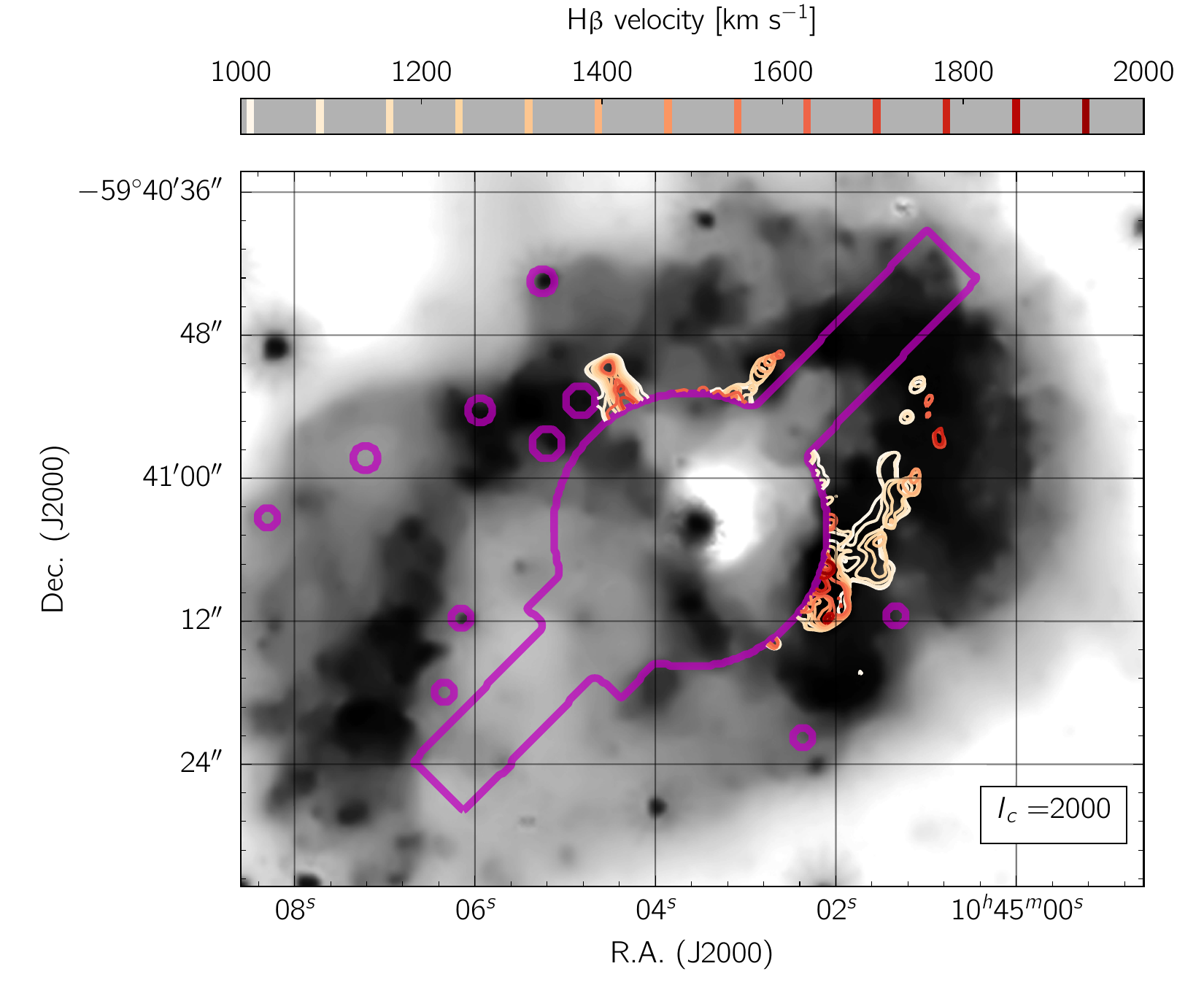}
\end{minipage}%
\begin{minipage}{.5\textwidth}
  \centering
  \includegraphics[width=1\linewidth]{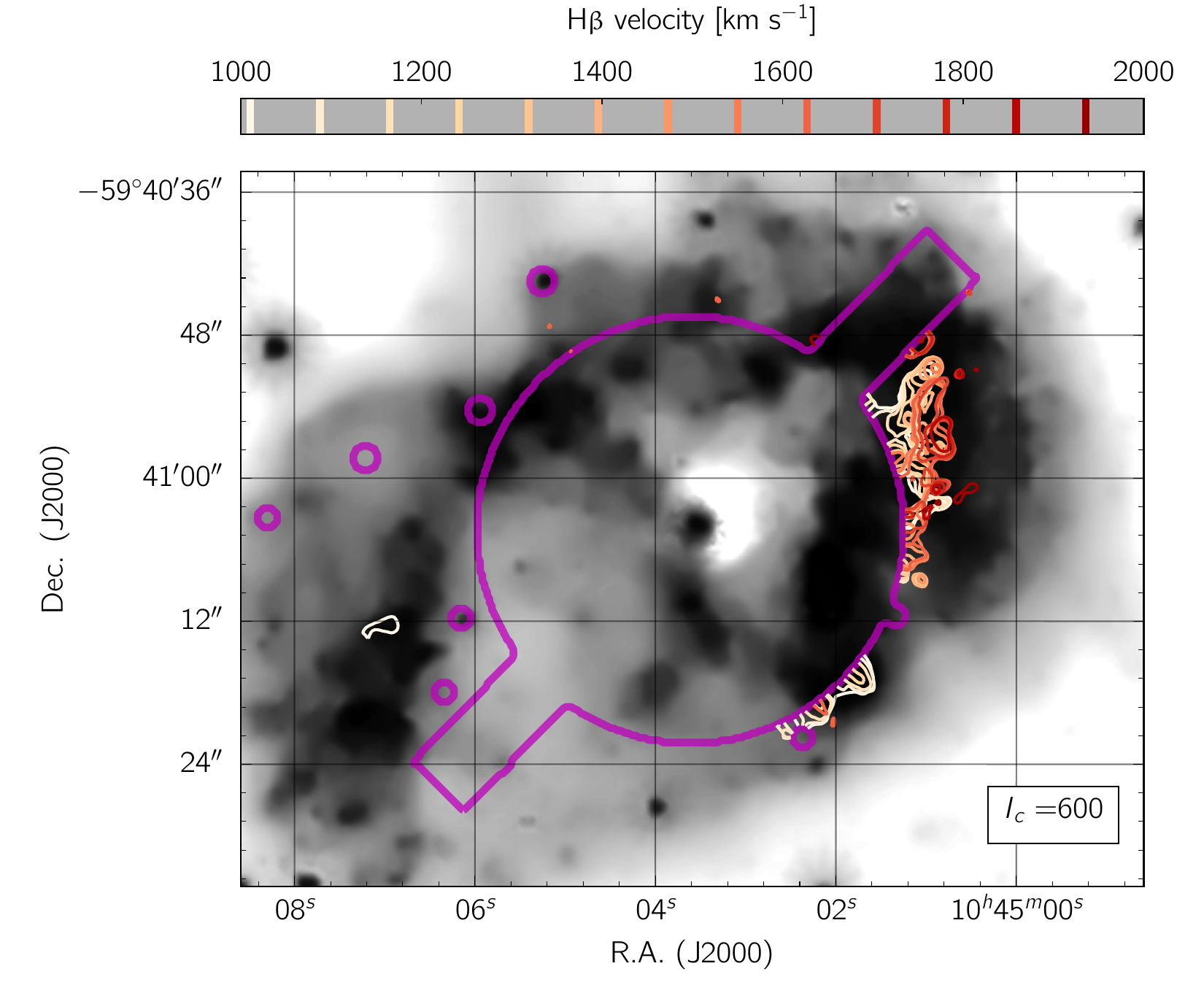}\end{minipage}
  \caption{The 0.5--1.2~keV {\it Chandra\/} X-ray image (gray-scale image) overlaid with the H$\beta$ velocity contours obtained from our MUSE data. Contours are fit in individual MUSE slices at an intensity level ``Ic''  in $10^{-20}$~erg/s/cm$^2$/\AA\ and color-coded as a function of velocity. The purple shapes indicate masked-out regions not used in this analysis that include the Homunculus, the artifacts from the optical ghosts described in Section \ref{data}, and several nearby stars.
Because of the number of artifacts present and the faintness of the clumps, two different mask-out regions are used for the positive velocity contours. We also adjust the ``Ic'' value in each case. Upper left: Extent of the soft X-ray emission surrounding $\eta$~Car. The green box indicates the region shown in the other three panels. Upper right: Contours of the blueshifted emission, highlighting the correlation of the optical emission with the soft X-rays in the southeast. The outer shell in the southeast coincides with the inner wall of the X-ray emission. Lower left and right: Contours of the redshifted emission. The red part of the outer shell is fainter and it is difficult to determine the velocity contours of the ejecta due to artifacts, but the optical emission here also traces well the inner wall of the X-ray emission.}
\label{fig:xray_vel}
\end{figure*}

X-ray images obtained with  {\it Einstein\/}, {\it ROSAT\/}, and {\it Chandra\/} show low-energy X-ray emission (0.2-–1.5~keV) in a hook shape around $\eta$~Car (Figure \ref{fig:xrays}; \citealt{1979ApJ...234L..55S,2001A&A...367..566W}). This soft X-ray emission is produced as ejected material runs into the surrounding gas and dust. Two brighter knots can be identified with the optical S ridge and the W arc, where also medium-energy X-ray emission (1.5–-3.0~keV) is observed.  There is very little soft X-ray emission in the southern part of the outer ejecta, where the density of optical clumps is also low. 
The central source and the Homunculus nebula are detected only in hard X-ray bands (3.0–-8.0~keV) with the emission from the Homunculus being reflected X-ray emission \citep{2004ApJ...613..381C}. No hard X-ray emission is detected outside the Homunculus.

The X-ray brightness is related to the combination of the speed of the ejecta and the density of the ambient medium. In the past, it was thus surprising that optical images showed only few direct correlations with the X-ray images, though optical emission was observed in most of the regions where X-ray emission is found \citep{2001A&A...367..566W,2004A&A...415..595W}. 
On the other hand, the expansion velocities of the ejecta were found to be in good agreement with the X-ray brightness distribution and X-ray temperatures, i.e., regions with higher X-ray emission have higher expansion velocities \citep{2001A&A...367..566W,2004A&A...415..595W}. The temperature of the X-ray gas is of the order of 0.65~keV, indicating velocities of the shocking gas on the order of 750~km~s$^{-1}$ \citep{2004A&A...415..595W}. This is consistent with the velocity structure seen in the MUSE data.  

The optical emission and velocity information from our MUSE data together with the SHAPE modeling let us predict where the X-ray emission (in three dimensions) should be -- and actually is. 
With the complete spatio-spectral sampling at optical wavelengths in the MUSE data we are able to recognize the excellent correlation between the X-ray emission and the optical ejecta. The large outer shell surrounding $\eta$~Car  aligns tightly inside the X-ray bubble and some bright X-ray features can be associated with counterparts in the optical (Figures \ref{fig:xrays} and \ref{fig:xray_vel}). 

Figure \ref{fig:xray_vel} shows the H$\beta$ intensity contours in the different MUSE cube wavelength slices, color-coded as a function of the corresponding line of sight velocity of the emitting gas. These contours are overlaid on top of the 0.5-1.2~keV {\it Chandra\/} X-ray image. The analysis is restricted to the regions outside the Homunculus due to the artifacts close to the central star in the final datacube, which are described in Section \ref{data}. The  figure shows the regions (purple shapes) that we mask out for this analysis. They contain the entire Homunculus, the concentric reflection rings around the central star, and several nearby stars. At negative velocities the intensity contours at different velocities trace clearly the expanding shell and the largest velocities align with the bright X-ray emission rim in the southeast. The bowshock is also distinctly resolved. The H$\beta$ emission at positive velocities is fainter and even very weak artifacts in the data make it difficult to determine their correlation with the X-rays. We thus chose two different mask-out region sizes  (Figure \ref{fig:xray_vel}, lower left and lower right) and different intensity levels to determine the H$\beta$ intensity contours at different velocities. We find that also for positive velocities the cylinder is clearly traced and the largest velocities align well with the bright X-ray hook.\footnote{Note that the artifacts are primarily concentric reflection rings around the central source and horizontal and vertical stripes, which could be easily identified as artifacts in this analysis.}

The NN bow is not associated with strong soft or medium X-ray emission, which may point to a high foreground column density \citep{2004A&A...415..595W}.
The X-ray bridge between the northern and southern part of the X-ray nebula could be attributed to a large expanding disk, maybe an extension of the equatorial disk, which produces shocks at the outer edge \citep{1995RMxAC...2...17D}. The lack of soft X-ray emission from the NN bow may suggest that this feature lies in front of the disk and absorbs the X-rays \citep{2004A&A...415..595W}. 
An alternative explanation for the absence of X-rays from the NN bow is that the material expands nearly freely into a low density region \citep{2004A&A...415..595W}.

\section{Discussion}
\label{discussion}

It is not yet resolved if $\eta$~Car's outer ejecta originate from eruptive events or high velocity winds prior to the Great Eruption, or if they were expelled during the Great Eruption. Eta~Car's history of sudden changes in its light curve and its spectrum, which have occurred quasi-periodically every $\sim$50~yr since the Great Eruption \citep{2008AJ....135.1249H,1999AJ....118.1777D} may indicate that this eruption was not an isolated event. It is possible that eruptive events have occurred in the past, which could be held responsible for the formation of the outer ejecta. Several (theoretical) models aim to explain the origin of the Homunculus (for references see Section \ref{intro}), but no working model exists yet for $\eta$~Car's outer ejecta. \citet{2004ApJ...616..976G} is the only example of a theoretical work which investigates if a high velocity pre-outburst wind just before the Great Eruption could have produced the ghost shell and the outer ejecta. Their work is inconclusive. 

To better understand the nature of $\eta$~Car's Great Eruption, it is essential to determine the three-dimensional kinematics and geometry of the ejected material. The unique value of the MUSE data is that they enable us to sketch the three-dimensional structure of $\eta$~Car's outer ejecta. With the help of the observed velocities, kinematic bridges can be built between morphologically isolated islands.

\subsection{The outer shell}

Eta Car's ejecta are chemically stratified, following a clear pattern of progressive nitrogen enrichment. Coincident with the soft X-ray shell are less nitrogen-rich ejecta than the material just outside the Homunculus nebula, and the gas beyond the X-ray shell has not been significantly processed through the CNO-burning cycle (\citealt{2004ApJ...605..854S}, see also \citealt{1997PASJ...49...85T,1998ApJ...494..381C,2007PASJ...59S.151H}). The outer ejecta thus interact with the unprocessed material from previous stellar wind mass loss. This suggests that $\eta$~Car has been ejecting nitrogen-enriched material only in the past few thousand years. 

Fast material in $\eta$~Car's outer ejecta, in excess of 1\,000~km~s$^{-1}$, has been discussed since the late 1980s (e.g., \citealt{1989RMxAA..18...87D,1996MNRAS.282.1313M,2001A&A...367..566W}). \citet{2008Natur.455..201S} estimated velocities of up to $3\,500$--$6\,000$~km~s$^{-1}$ and proposed that these high velocities indicate that the material has been accelerated by the pressure behind a ``blast wave'', which originated from a deep-seated explosion. This ``blast wave'' involves only a small amount of high-velocity material originating from $\eta$~Car's SN impostor event and should not be confused with the blast waves seen in SN explosions.
\citet{2012Natur.482..375R} derived from light-echo spectra that $\eta$~Car was very cool during the Great Eruption with an effective temperature of $\sim$5\,000~K and argued that this supports the notion that a physical mechanism such as an energetic blast wave may be associated with the outer ejecta \citep{2008Natur.455..201S}, but see \citet{2012Natur.486E...1D}.
The Great Eruption may thus have been powered by a deep-seated explosion. Candidates for an explosion mechanism could be a pulsational pair instability \citep{2007Natur.450..390W} or another instability associated with nuclear burning in the last stages of evolution. These events are expected to occur only $\sim$10--1\,000~yr before the final core collapse SN, as proposed for the progenitors of the extremely luminous SN 2006gy \citep{2007ApJ...666.1116S} and SN 2006jc \citep{2007Natur.447..829P}.

Some models for $\eta$~Car's Great Eruption, however, assume a radiation-driven super-Eddington wind \citep{1994PASP..106.1025H,1997ARA&A..35....1D,2000ApJ...532L.137S,2004ApJ...616..525O}. A recent review by \citet{2016JPhCS.728b2008D} advocates strongly that LBV giant eruptions are super-Eddington mass outflows. 
\citet{2016MNRAS.462..345O} found that the low temperatures from light echoes of $\eta$~Car's Great Eruption are consistent with the very large mass-loss rates and luminosities estimated for this eruption epoch.
However, \citet{2016MNRAS.462..345O} also note that such a cool spectral temperature is not unique to a steady wind outflow model and cannot be used by itself to discriminate between the explosion and steady-wind scenarios.

As explained in Section \ref{shape}, we can only roughly constrain the age of the outer shell surrounding $\eta$~Car with the SHAPE modeling, because of the degeneracy between the velocity field and the linear distance of ejecta from the central star. 
Without any historical record of eruptions prior to the Great Eruption and assuming that material decelerates, it is most appropriate (for all intents and purposes) to presume that all outer shell material was ejected by the energy release that caused the Great Eruption. (Note, that observations in the 18th century were sparse and we cannot entirely rule out eruptions prior to the Great Eruption, see Section \ref{shape}.) The shell has a thickness of approximately 15\% of its radius, which implies an upper limit of 25~yr for its formation and is thus consistent with the formation during the Great Eruption (e.g., \citealt{2011MNRAS.415.2009S}).
The fact that the outer ejecta have a slightly different orientation than the Homunculus of $\sim$10\degree\ does not indicate ejection in different directions and at different times. The cylindrical structure is also not perfectly straight and the orientation varies locally by a similar amount. Variations in the orientation could be due to interactions with the inhomogeneous environment. Without further study, the significance of these deviations cannot be assessed. 
Slower knots and arcs surrounding $\eta$~Car, such as the E condensations, may be older, consistent with their less nitrogen-enriched chemical abundances.

\citet{2008Natur.455..201S} proposed a bipolar forward shock geometry similar to the Homunculus, but three to four times its size and speed for the shape of the outer ejecta. This would result in an up-scaled version of the Homunculus, yet our data are inconsistent with this idea. We find that the outer shell is not a scaled version of the Homunculus nebula. If we assume the same bipolar geometry as for the Homunculus, we cannot match the observed spatial distribution of the outer ejecta with a SHAPE model (Appendix \ref{sec:alternative}). The geometry we derive is instead more appropriately described by a bent tube (Figure \ref{fig:largescale}). \citet{2008Natur.455..201S} suggested that the interaction between the ``blast wave'' from $\eta$~Car's Great Eruption with 500--1\,000~yr old clumpy ejecta gives rise to the soft X-ray emission. We confirm a strong correlation between the outer shell and the soft X-ray emission with our MUSE data (Figures \ref{fig:xrays} and \ref{fig:xray_vel}, see also \citealt{2001A&A...367..566W}). The outer shell tightly aligns inside the X-ray bubble and bright optical emission is associated with bright X-ray features.

We revise the ejecta velocities in \citet{2008Natur.455..201S}, who estimated a range for the largest observed (deprojected) velocities in $\eta$~Car's outer ejecta between 3\,500 and 6\,000 ~km~s$^{-1}$, depending on the assumed geometry. From the derived three-dimensional geometry of the outer ejecta with the MUSE data and SHAPE modeling, we find that the largest deprojected velocities are about 4\,500~km~s$^{-1}$.

\subsection{The ghost shell}

\citet{2002A&A...389L..65C} proposed that the ghost shell in front of the southern Homunculus lobe is material associated with the forward shock between the fast stellar wind of the Great Eruption and the older, slower, massive wind. The reverse shock formed the leading edge of the Homunculus. This would be consistent with the lack of soft X-rays from the Homunculus, since the material would be expanding into a low density medium. 

We find that the ghost shell consists of two aligned shells in front of the southern Homunculus lobe. The shells expand with velocities of only a few hundred km~s$^{-1}$ faster than the Homunculus material. A scenario that could accommodate this invokes several pulses of ejection during the Great Eruption (maybe a few years apart), during which material was expelled, forming the two ghost shells and the Homunculus nebula, but possibly also involving a later shock/reverse shock velocity separation.

\subsection{The NN bow}
\label{ageNN}

The NN condensation and the associated NN bow consist of fast moving material (deprojected velocities are about 1\,300~km~s$^{-1}$, the deprojected extent from the star is about 0.22~pc) close to the plane of the equatorial skirt. Together with the measured proper motion between the 1997 {\it HST\/} WFPC2 images and the 2014 MUSE data, we estimate an ejection date about 170 years ago, which (again) points to the Great Eruption. This suggests that the main structures in the ejecta surrounding $\eta$~Car, i.e., the Homunculus, the ghost shell, the NN bow, and the outer shell are all product of the Great Eruption.
It is not clear what mechanism could have ejected material with these velocities in a direction that is perpendicular to the Homunculus axis and located in the presumed orbital plane. 

We have shown that a hydrodynamical model for a jet or ``bullet'' reproduces the morphology of the NN bow well. However, this does not mean that this feature is a jet, as suggested by \citet{1993A&A...276L..21M}.
There is no consensus on the mechanism for forming jets from eruptive stellar systems. For $\eta$~Car, the interacting companion offers a plausible cause. The jet could have originated from an outflow following the accretion of material by the secondary star \citep{2005ApJ...635..540S,2013NewA...18...23A,2016RAA....16e...1S}. In this case, however, a peculiar orientation of the accretion disk needs to be invoked. There might have been an explosive instability in the accretion disk around the secondary star.  
In this case, the lack of a counter jet is also more readily understood. 

The NN bow resembles a slingshot prominence, i.e, a large arc or loop. It is somewhat reminiscent of structures that can be found in the ejecta of the red supergiant VY~CMa and the post-red supergiant IRC+10420 (e.g., \citealt{2007AJ....133.2716H,2010AJ....140..339T}). In these objects the eruptive mass loss is probably due to convective activity and magnetic fields. 
However, while the NN bow has a similar morphology and spatial extent as the arcs seen in these objects, the observed velocities are 1--2 orders of magnitude larger. Eta~Car is also at least an order of magnitude more luminous than VY CMa and IRC+10420. In contrast to $\eta$~Car, the large arcs in VY~CMa were ejected in different directions over several hundred years; in IRC+10420 the semi-circular arcs are equatorial. These objects are clearly very different from $\eta$~Car, but perhaps the NN bow represents a counterpart of the arc-like, episodic eruptive events seen in these less luminous objects.  The similar effective temperature reached by $\eta$~Car during the Great Eruption (though not in hydrostatic equilibrium) may suggest that a similar physical mechanism is at play in both cases, and further theoretical work is warranted.

One could also imagine more exotic explanations for the NN bow. For example, \citet{2016MNRAS.456.3401P} discuss the Great Eruption in terms of a merger event in a triple system, with massive loops of ejected material.
In their model, the Homunculus was produced prior to the merger by an extremely enhanced stellar wind, energized by tidal energy dissipation. The merger itself resulted in a massive asymmetric outflow (two loops) in the equatorial plane.

\section{Conclusion}
\label{conclusion}

Three-dimensional information on the morphology and kinematics of $\eta$~Car's ejecta provides evidence for their ejection history and structure. 
In the past, $\eta$~Car's outer ejecta were seen as a conglomerate of individual small structures. This was regarded as evidence for very rapid and efficient fragmentation within the nebula or a mechanism which favors the ejection of clumps. 
Using MUSE IFU observations and the modeling tool SHAPE, we reconstructed the large-scale three-dimensional geometry of $\eta$~Car's outer ejecta and show their contiguous nature.

Our analysis suggests that the star is surrounded by a large bent partial cylinder (the outer shell), centered on the star and roughly aligned with the Homunculus. This structure fits tightly into the X-ray bubble. For the modeling with SHAPE, we have to make an assumption on the velocity field and on the ejection times of the ejecta. Based on several considerations, described in Sections \ref{shape} and \ref{ageNN}, the probable ejection time of the outer shell is during the Great Eruption and it is improbable that the expelled material is more than twice as old as the Great Eruption. 

The existence of two ``ghost'' shells just outside the southern Homunculus lobe, and the NN outflow hints at a sequence of several outbursts during the Great Eruption, and/or later shock/reverse shock velocity separation of material. This confirms the notion of event-driven mass loss, breaks it down to smaller scales, and poses an invitation for theorists to search for an explanation. 

In future work, it will be interesting to study in detail the correlation between the variations in X-ray temperature and the radial velocities of the ejecta. The MUSE data also provide for detailed velocity and abundance studies, which will shed further light on the chemical composition during different ejection episodes. However, the blending of lines due to the large velocity ranges of ejecta within the same line of sight makes this a difficult task.
The MUSE data set and the SHAPE modeling show that three-dimensional geometry is needed to study $\eta$~Car's outer ejecta, against which future three-dimensional hydrodynamical and radiative transfer simulations should be compared to. 

\vspace{1\baselineskip}\vspace{-\parskip}
{\it Note.} During the second revision of this paper, \citet{2016MNRAS.463..845K} was published, whose content is directly related to our paper. \citet{2016MNRAS.463..845K} measured proper motions of about 800 features in $\eta$~Car's outer ejecta with a baseline of over 21~yr. They find that the velocities are consistent with features moving at constant velocities, as we have assumed in our modeling. The authors argue for two ejection dates for the outer ejecta prior to the Great Eruption: 
\begin{enumerate}
\item 
In the mid-1200s: the E and NNE condensation originated in this event. These features are not part of our model and we agree that they are from a previous mass loss episode. 
\item
An intermediate eruption in the 16th century: the SE arc, the W condensation, and the NW condensation originated in this event. These three features are also not part of our model. They are certainly distinct to the features that comprise the coherent outer shell structure, which originated in the Great Eruption. However, we assume that the S ridge is associated to the Great Eruption.
\end{enumerate} 

\begin{appendix}
\section{A resized Homunculus model for the outer shell}
\label{sec:alternative}

\begin{figure*}[p]
\centering
  \includegraphics[width=1\linewidth]{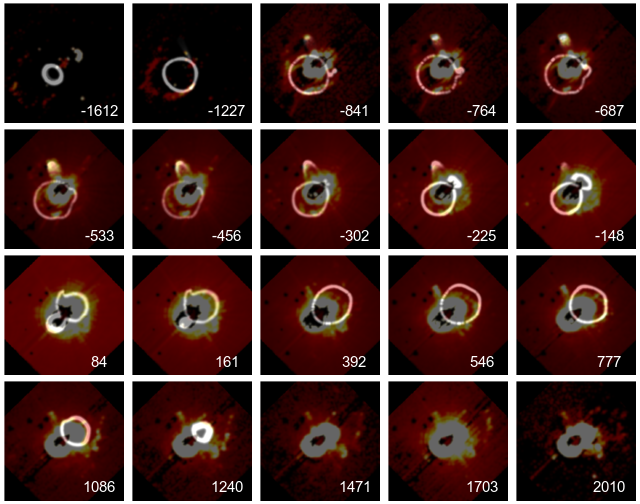}
  \caption{Alternative SHAPE model of a resized Homunculus geometry compared to the observed MUSE channel maps of H$\beta$ emission at different velocities in the MUSE field of view of 1\arcmin$\times$1\arcmin. This geometry assumption does not reproduce the observations nearly as well as the cylinder geometry shown in Figure \ref{fig:channelmaps}. With a given velocity field, there is not much freedom in the geometry.}
\label{fig:alternatives_3}
\end{figure*}

Figure \ref{fig:alternatives_3} shows a SHAPE model of a resized Homunculus geometry for the outer ejecta. This geometry assumption does not reproduce the observations nearly as well as the cylinder geometry shown in Figure \ref{fig:channelmaps}.  By altering the scale and the speed (i.e., age) of the resized Homunculus model, the match with the observational data becomes only worse than the case shown here.

\end{appendix}

\begin{acknowledgements}  
This research has made use of NASA's Astrophysics Data System Bibliographic Services, \textsc{SAOImage DS9} \citep{2003ASPC..295..489J}, developed by Smithsonian Astrophysical Observatory, QFitsView (\url{http://www.mpe.mpg.de/~ott/dpuser/qfitsview.html}), \textsc{Aladin sky atlas} developed at CDS, Strasbourg Observatory, France \citep{Bonnarel2000,2014ASPC..485..277B}, \textsc{Astropy}, a community-developed core Python package for Astronomy \citep{AstropyCollaboration2013}, \textsc{APLpy}, an open-source plotting package for Python hosted at \url{http://aplpy.github.com}, and of \textsc{Montage}, funded by the National Science Foundation under Grant Number ACI-1440620, and was previously funded by the National Aeronautics and Space Administration's Earth Science Technology Office, Computation Technologies Project, under Cooperative Agreement Number NCC5-626 between NASA and the California Institute of Technology. 
W.S.\ acknowledges support from grant IN101014 from UNAM-DGAPA-PAPIIT.
\end{acknowledgements}

\bibliographystyle{aa}

\end{document}